\documentclass[twocolumn,english,prx]{revtex4-1}
\usepackage{color}
\usepackage[utf8]{inputenc}
\usepackage{graphicx}
\usepackage[T1]{fontenc}
\usepackage{float}
\usepackage{natbib}
\usepackage{textcomp}
\usepackage{amsmath}
\usepackage{amssymb}
\usepackage{graphicx}
\usepackage{esint}
\usepackage{natbib}
\usepackage{xcolor}
\usepackage{multirow}
\usepackage{ulem}

\begin{document}
\title{Electron interferometry and quantum spin Hall phase in silicene}

\author{Bart\l{}omiej Rzeszotarski}

\affiliation{AGH University of Science and Technology, Faculty of Physics and
Applied Computer Science,\\
 al. Mickiewicza 30, 30-059 Kraków, Poland}

\author{Alina Mre\'n{}ca-Kolasi\'n{}ska}

\affiliation{AGH University of Science and Technology, Faculty of Physics and
Applied Computer Science,\\
 al. Mickiewicza 30, 30-059 Kraków, Poland}

\author{Bart\l{}omiej Szafran}

\affiliation{AGH University of Science and Technology, Faculty of Physics and
Applied Computer Science,\\
 al. Mickiewicza 30, 30-059 Kraków, Poland}

\begin{abstract}
We discuss devices for detection of the topological insulator phase based on the two-path electron interference. For that purpose we consider buckled silicene for which a local energy gap can be opened by vertical electric field to close one of the paths and for which the quantum spin Hall insulator conditions are controlled by the Fermi energy. In quantum spin Hall phase the interference is absent due to the separation of the spin currents and
the conductance of the devices include sharp features related to localized resonances. In the normal transport conditions the two-path interference produces a regular Aharonov-Bohm oscillations in the external magnetic field.
\end{abstract}
\maketitle

Quantum spin Hall (QSH) insulators \cite{r1, r2, r3}  form a class of two-dimensional topological insulators
with bulk energy gap and topologically protected  currents of a fixed spin-orbital helicity. The QSH phase \cite{r4} is discussed for bulk nanostructures including HgTe quantum wells \cite{r6, r7, r8, r9} and InAs/GaSb interfaces \cite{r10,r11} as well as graphene-like monolayer Xenes materials \cite{r12,r13}, including silicene \cite{silitmdc,chow,Liu11,Liu,Ezawa}.
 The  QSH conditions in silicene  occur for the Fermi energy  near the charge neutrality point \cite{silitmdc,chow,Liu11,Liu,Ezawa}. The Fermi energy in 2D monolayer materials can be controlled by external gating. 
In the QSH phase the spin currents  are confined by opposite edges of the sample,
which was used for proposals of spin sources and spin filters in silicene \cite{r28,r29,r39,r40,r42,r33}. 
In this paper we propose electron interferometer devices that can be used for
detection of the QSH transport conditions. The devices are based on the idea of two-path interference
and the spin separation by the split silicene ribbon \cite{r28}. 
We consider a double slit interference device as well as a quantum ring and find that
in the normal phase one observes smooth Aharonov-Bohm conductance oscillations 
while in the QSH regime only sharp conductance features due to the localized resonances 
with circular current loops are observed.
In silicene both the localized resonances and the Aharonov-Bohm oscillations can be 
intentionally switched off by applying a local electric field to one of the arms
of the split channels, due to the buckling of the crystal lattice that translates
the electric field into a local energy gap \cite{fie1,fie2} that stops the current flow.

\begin{figure}[htbp]
\centering
\includegraphics[width=0.3\textwidth]{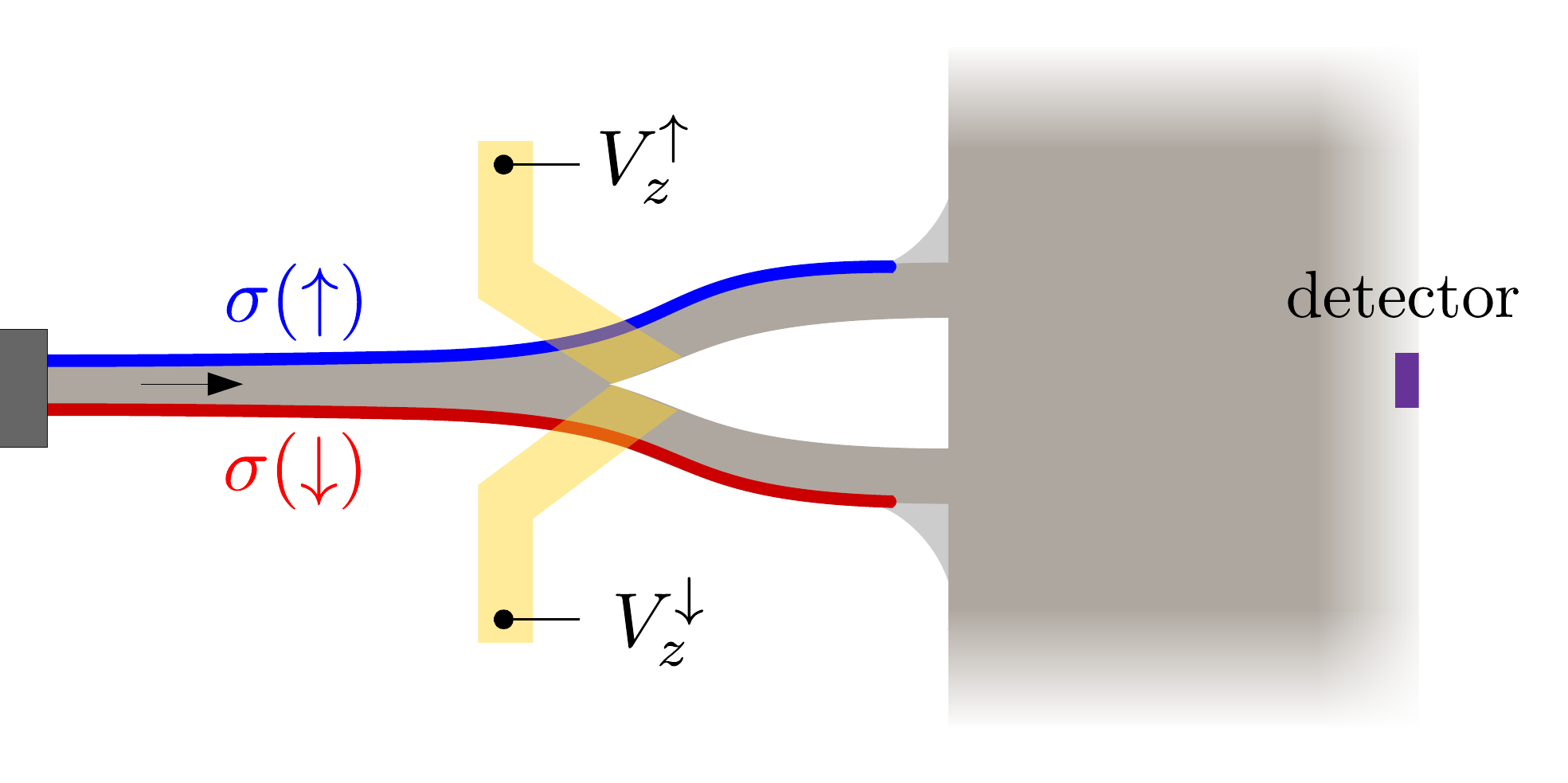}\includegraphics[width=0.2\textwidth]{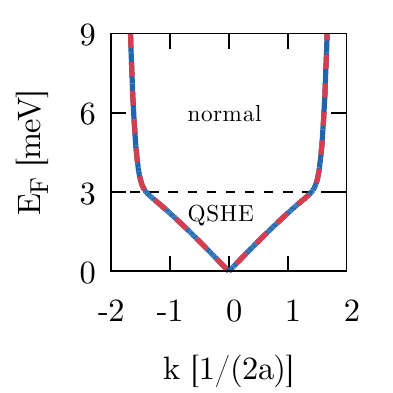}
\put(-245,100){(a)}\put(-100,100){(b)}
\caption{(a) Sketch of the split channel  system. The input lead and
the split channels are silicene  ribbons 
of width 6.5 nm and zigzag edges.
In the quantum spin Hall effect (QSHE) phase each channel is fed by different spin-state current from input lead (the blue and red lines at the edges of the channel). 
The length of the split part is about 60 nm and the vertical spacing
between the split channels is 7 nm. The horizontal distance between openings of the slits
and the detector is 32 nm. The detector is a ribbon 6.5 nm wide. External gates marked in yellow can be used to locally open the energy gap in silicene.
(b) The spin-degenerate dispersion relation of the zigzag 6.5 nm wide silicene ribbon
at the conduction band side. The linear energy range corresponds to the 
QSH insulator phase. For higher $E_F$ the spin for both spin orientations flows through the center of the ribbon.
}
\label{fig:sch1}
\end{figure}



The schematics of the double slit interferometer depicted in [Fig. \ref{fig:sch1}(a)]. 
The electrons are fed from the left by the silicene ribbon of a zigzag edge
of 6.5 nm width.  The zigzag ribbon
 supports the spin-polarized edge transport at the  Fermi energy $E_F\in(-3, 3)$ meV with respect to the charge neutrality point [see the dispersion relation in Fig. \ref{fig:sch1}(b)].  In the quantum spin Hall insulator phase  the opposite spin currents flow at the opposite
edges of the ribbon [see Fig. \ref{fig:sch1}(a)]. 
The input lead  splits into two channels of the same width.
In the topological phase this spindle-shaped connection separates the opposite spin currents 
to the two channels [Fig. \ref{fig:sch1}(a)].
The split  channels are connected to semi-infinite open plane of silicene [Fig. \ref{fig:sch1}(a)] with a smoothed extensions that prevent backscattering. 
At open halfplane the areas marked by the gray color
fading to white in Fig. \ref{fig:sch1}(a) we attach wide silicene ribbons that make the edges of the computational box  reflectionless. 
For $E_F>3$ meV [Fig. \ref{fig:sch1}(b)] the current flows through the bulk of channel for both spin orientations and we refer to these conditions as the normal phase. 
In the normal phase the  current flows through both the split channel for both spin orientations. The Young interference of the waves entering the open halfplane by different slits can only occur in the normal phase.  In the topological phase each of 
the slits feeds opposite spin. Thus, observation of the Young interference should 
depend on the Fermi energy.
In order to monitor the interference in the model device [Fig. \ref{fig:sch1}] at 32 nm to  the right of the slit opening a zigzag ribbon  of width 6.5 nm is connected as a detector \cite{kka}.
In order to gain additional control in the interference device we introduce local
gates (yellow gates in Fig. \ref{fig:sch1}) to switch off the currents.

\begin{figure}[htbp]
\centering
\includegraphics[width=0.45\textwidth]{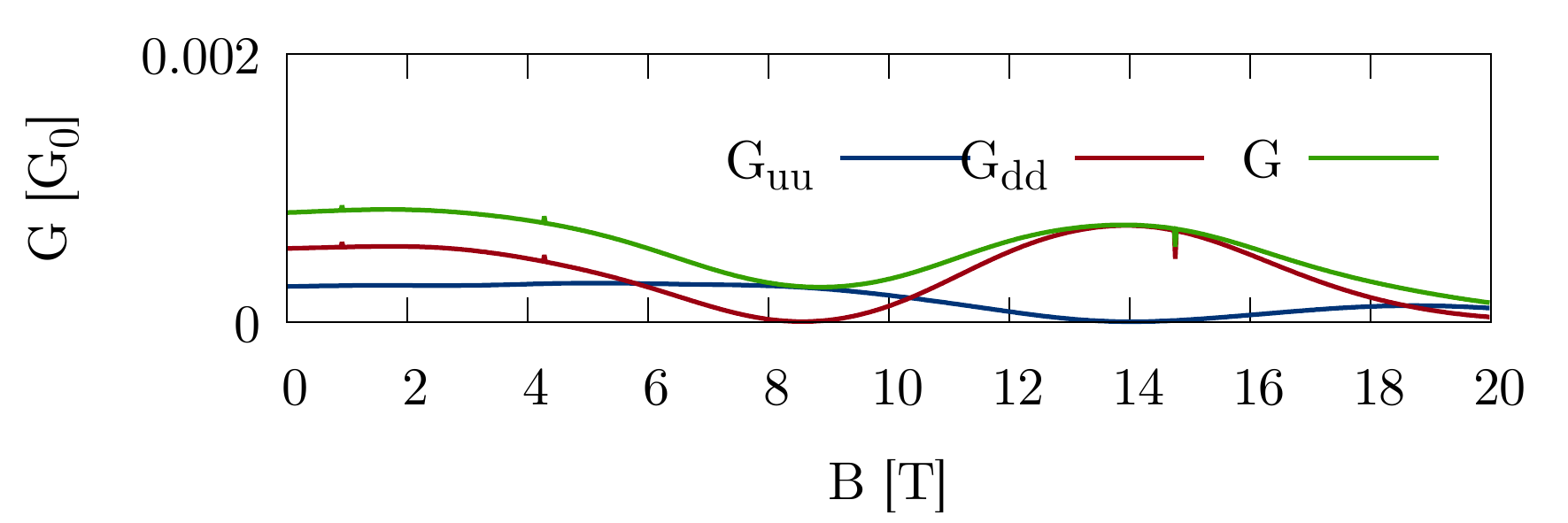}\put(-225,65){(a)}\put(-180,55){$V_z^{\uparrow}=0$}

\includegraphics[width=0.45\textwidth]{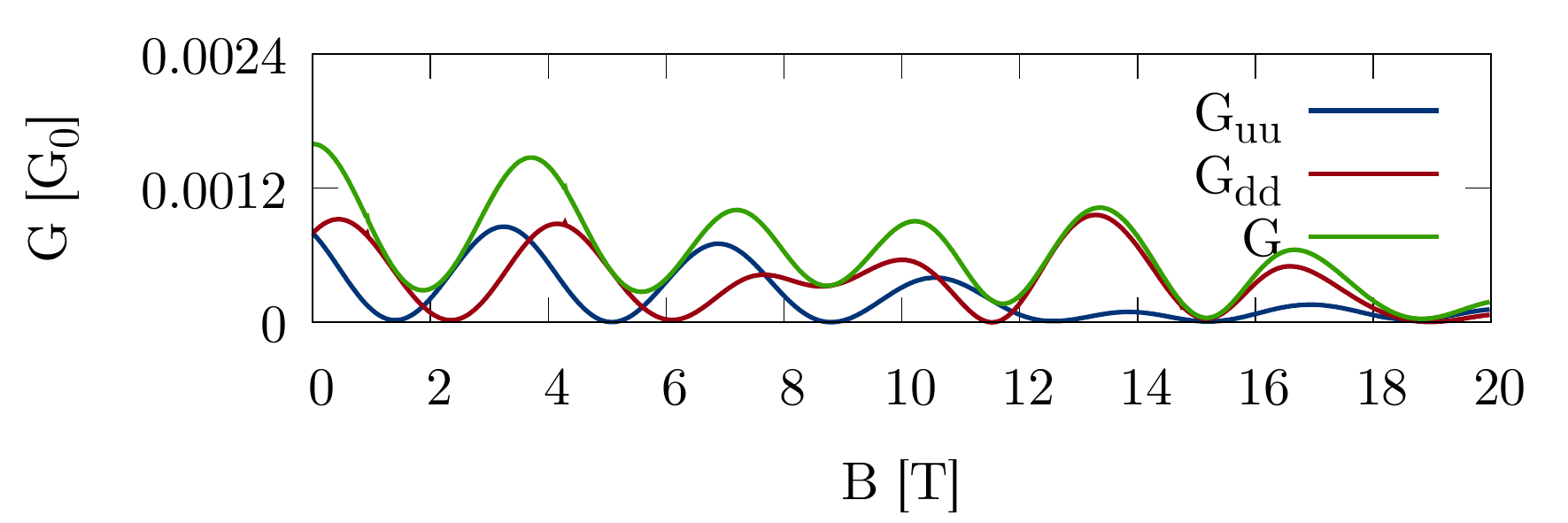}\put(-225,65){(b)}\put(-125,50){$E_F$ = 5meV}

\includegraphics[width=0.45\textwidth]{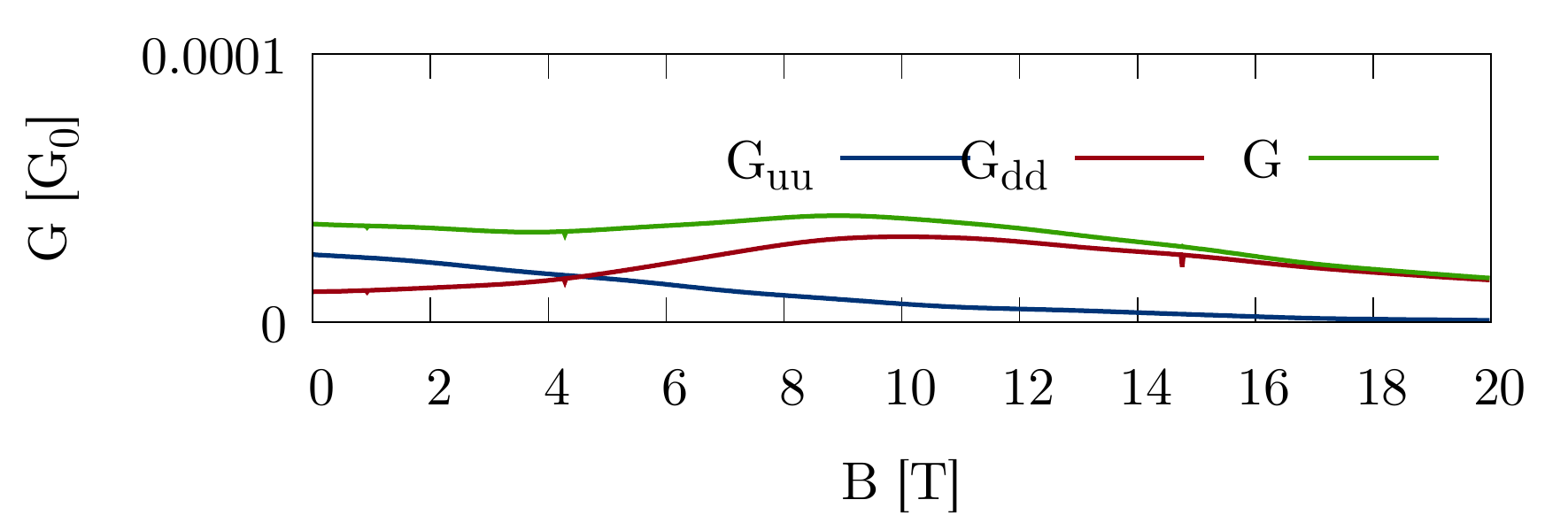}\put(-225,65){(c)}\put(-180,55){$V_z^{\downarrow}=0$}
\caption{ Conductance with the upper (a) lower (c) and both (c) channels open for $E_F$ = 5meV (outside the QSHE regime). In (a) [(c)] the vertical electric field 100 meV/\AA is applied to the lower [upper] channel.}
\label{fig:cut}
\end{figure}

\begin{figure}[htbp]
\centering
\includegraphics[width=0.45\textwidth]{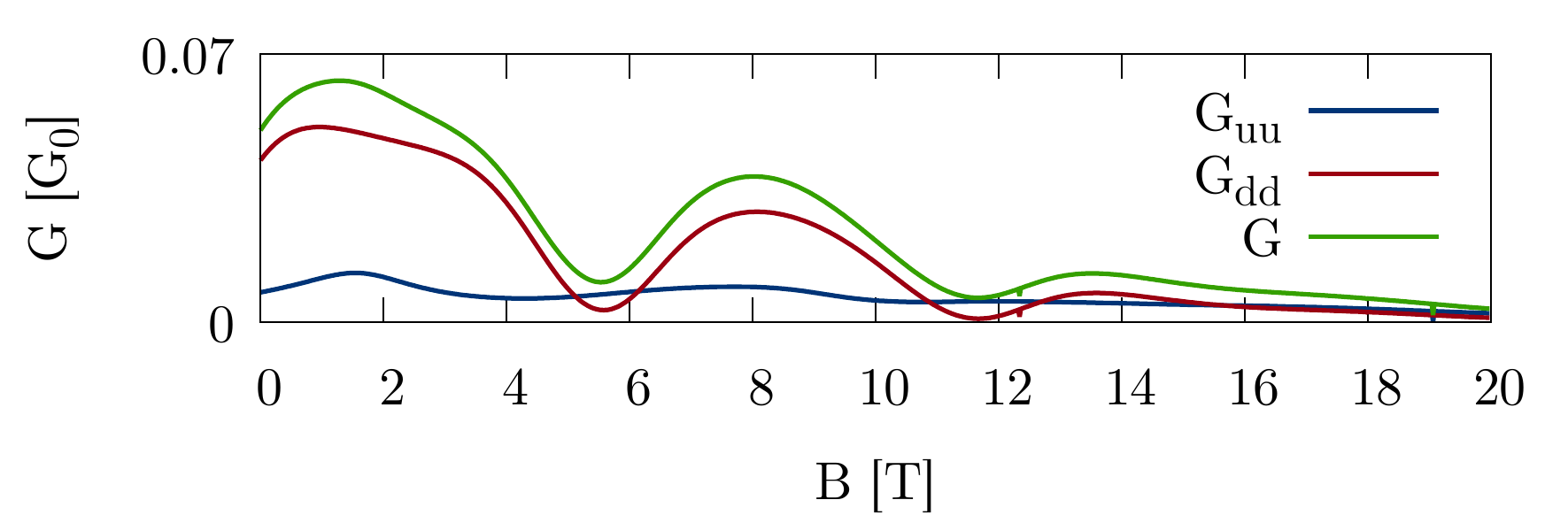}\put(-225,65){(a)}\put(-135,55){$V_z^{\uparrow}=0$}

\includegraphics[width=0.45\textwidth]{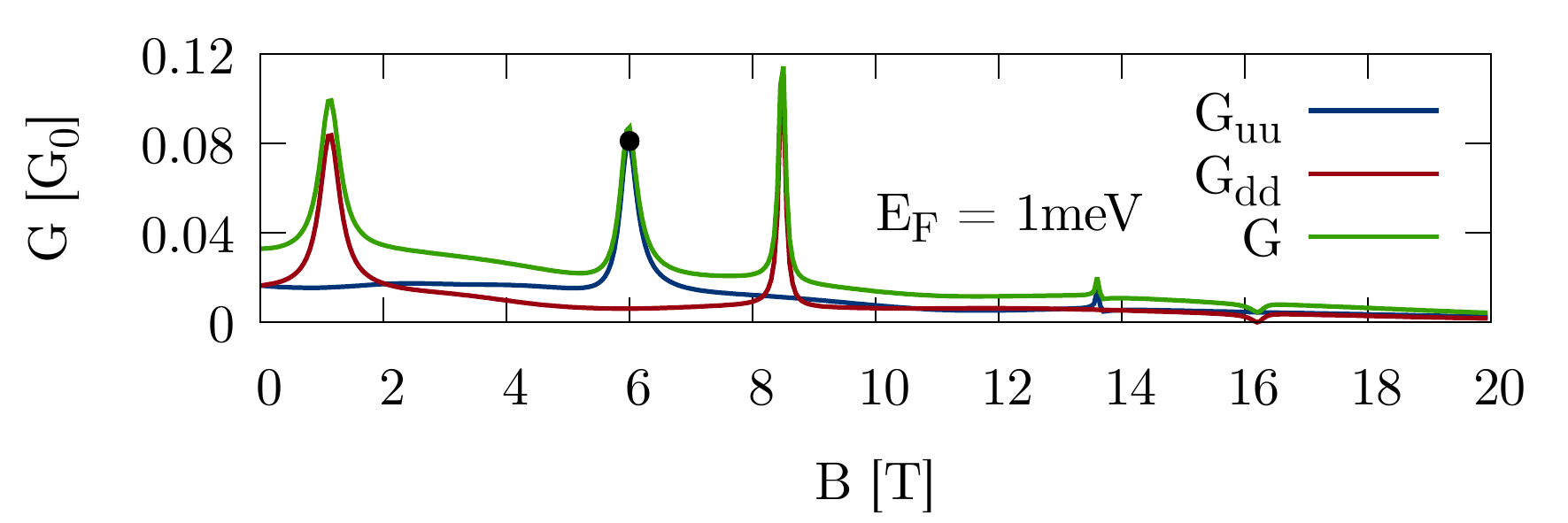}\put(-225,65){(b)}

\includegraphics[width=0.45\textwidth]{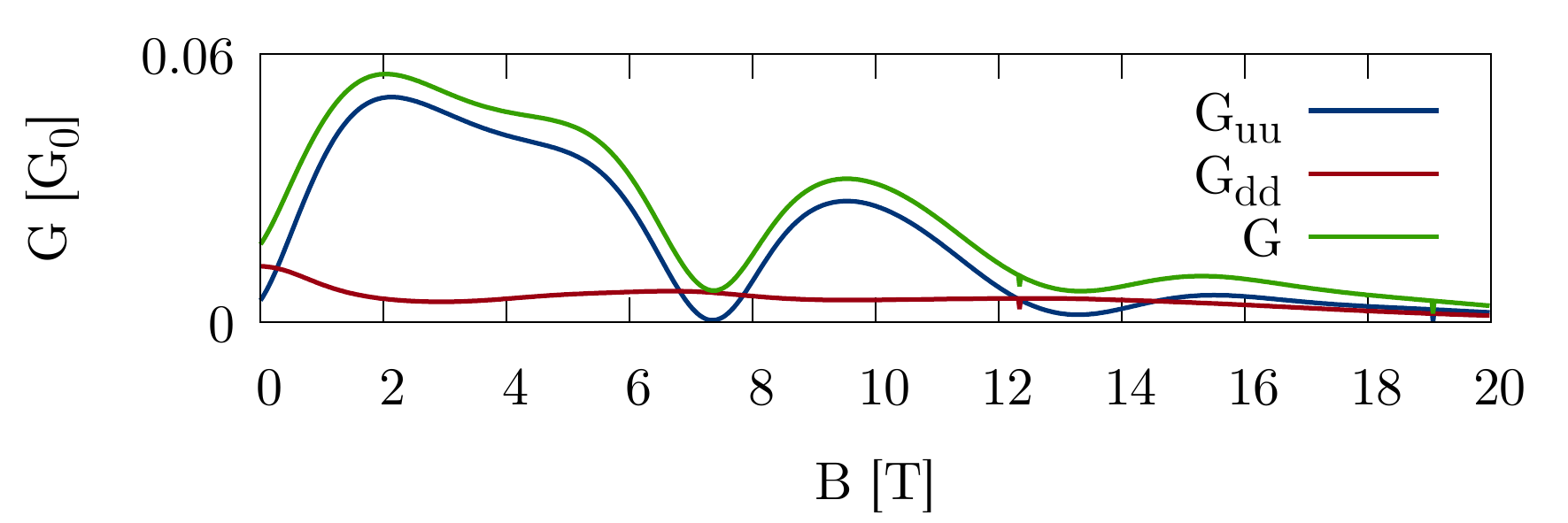}\put(-225,65){(c)}\put(-135,55){$V_z^{\downarrow}=0$}
\caption{  (b) Two-slit interference for $E_F$ = 1meV (in the QSHE regime) in comparison to the one-slit transmissions (a,c). In (a) upper slit is open ($V_z^{\uparrow}=0$ and $V_z^{\downarrow}=100$meV/\AA) while in (c) lower slit is open ($V_z^{\uparrow}=100$meV/\AA{} and $V_z^{\downarrow}=0$). }
\label{fig:cutq}
\end{figure}

We use  the tight-binding Hamiltonian spanned on $p_z$ orbitals of Si atoms \cite{Liu} 
\begin{align}
H=&-t\sum_{\langle i,j\rangle \alpha }  c_{i \alpha}^\dagger c_{j \alpha}+i\frac{\lambda_{SO}}{3\sqrt{3}} \sum_{\langle \langle i,j\rangle \rangle \alpha, \beta } \nu_{ij} c^\dagger_{i\alpha} \sigma^{z}_{\alpha,\beta}c_{j\beta},  \nonumber \\ 
& -i\frac{2}{3}\lambda_{R}^{int.} \sum_{\langle \langle  i,j \rangle \rangle \alpha,\beta } \mu_{ij} c^\dagger _{i\alpha}\left(\vec{\sigma}\times\vec{d}_{ij} \right)^z_{\alpha\beta} c_{j\beta} \nonumber \\& +\sum_{i \alpha }  V_{zi} c_{i \alpha}^\dagger c_{i \alpha}
\label{eq:h0}
\end{align}
\noindent where $c_{i \alpha}^\dagger$ ($c_{j \alpha}$) is the creation (annihilation) operator for an electron on atom $i$ with spin $\alpha$. 
The calculation accounts for hexagonal lattice of Si atoms with constant 
$a=3.89$ \AA\; and a vertical shift of 0.46 \AA\; between the A and B sublattices. 
Summations over $\langle i,j\rangle$ and $\langle\langle i,j\rangle\rangle $ run over  nearest and next nearest neighbor ions, respectively. In Eq. (1) we use $t=1.6$ eV for the hopping energy \cite{Liu,Ezawa},   $\lambda_{SO}=3.9$ meV \cite{Liu} is the intrinsic spin-orbit coupling energy \cite{r4} where $\nu_{ij}=-1$  ($+1$) for the clockwise (counterclockwise) next-nearest neighbor hopping, $\lambda_{R}^{int.}=0.7$ meV is the intrinsic spin-orbit coupling energy \cite{Liu,Ezawa}, where the unit vector from $j$-th to $i$-th ion ${\bf d}_{ij}=\frac{{\bf r}_j-{\bf r_i}}{|{\bf r}_j-{\bf r_i}|}$. 
 Within sublattice A (B) we apply $\mu_{ij}=+1$ ($-1$).
The last term in Eq. (2) introduces a local vertical electric field
(yellow gates in Fig. \ref{fig:sch1}) at the entrance to the slits
 to intentionally switch off the currents in the channels.
The gates introduce vertical electric field of about 100 mV/\AA \;that produces the potential difference $\pm 25$ meV at the A and B sublattices of the buckled silicene lattice. The field opens the local energy gap for the Fermi energy range considered here
and closes the channel for the electron flow. 
In presence of an external perpendicular magnetic field $B_z$ the Peierls phase
is introduced to the hopping terms.


We derive the conductance of the device by solution of the quantum scattering problem
within the Landauer approach.
For the latter we use the wave function matching  \cite{r33,bubel} method for the atomistic description of the medium. The positive (negative) $\langle \sigma_z \rangle$ values are labeled by $u$,$\uparrow$ ($d$,$\downarrow$). The Rashba interaction is weak, the spins are nearly polarized in the $z$ direction and no spin flips are obtained ($G_{ud}=G_{du}=0)$ and the total conductance is a sum of spin-diagonal contributions $G=G_{uu}+G_{dd}$.

 Figure \ref{fig:cut}(b) shows the conductance in the absence of the electric field in the gated area  in the normal phase for $E_F=$ 5 meV.   The conductance for both
spin orientations and the  total conductance undergo periodic oscillations
in the external magnetic field with a period of about 4 T, which 
is the Aharonov-Bohm period for the area of about 1000 nm$^2$ enclosed between the electron paths passing through the split channels to the detector.
For the upper [Fig. \ref{fig:cut}(c)] or lower [Fig. \ref{fig:cut}(a)] channel cut off
 by the vertical electric field, the Aharonov-Bohm conductance oscillations disappear,
which is a signature of the switched off two-slit interference. 

  \begin{figure}[htbp]
\centering
\includegraphics[width=0.41\textwidth,clip,trim=0.2cm 1.8cm 0.5cm 1.3cm]{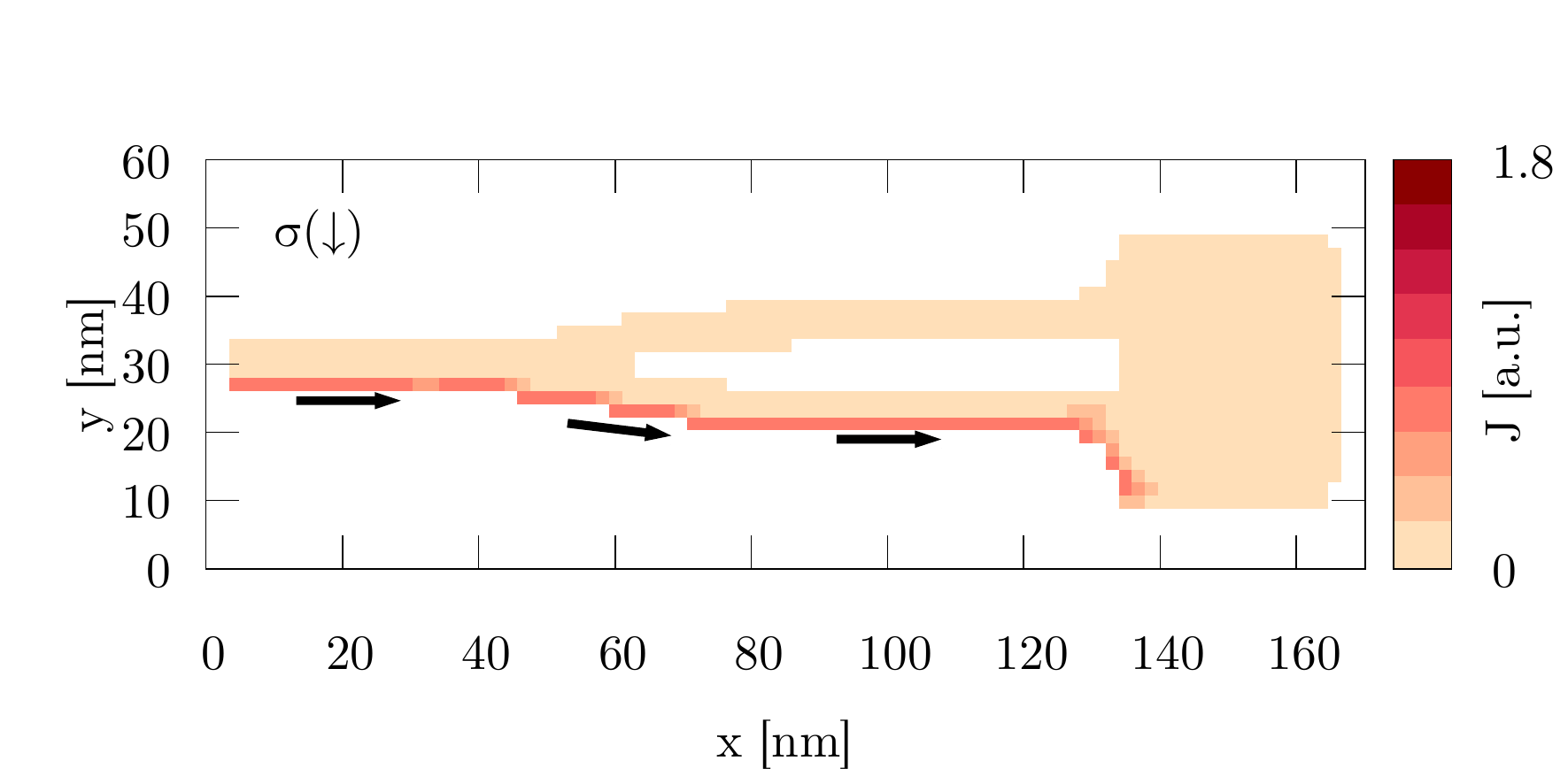}\put(-180,10){(a)}\\
\includegraphics[width=0.41\textwidth,clip,trim=0.2cm 1.0cm 0.5cm 1.3cm]{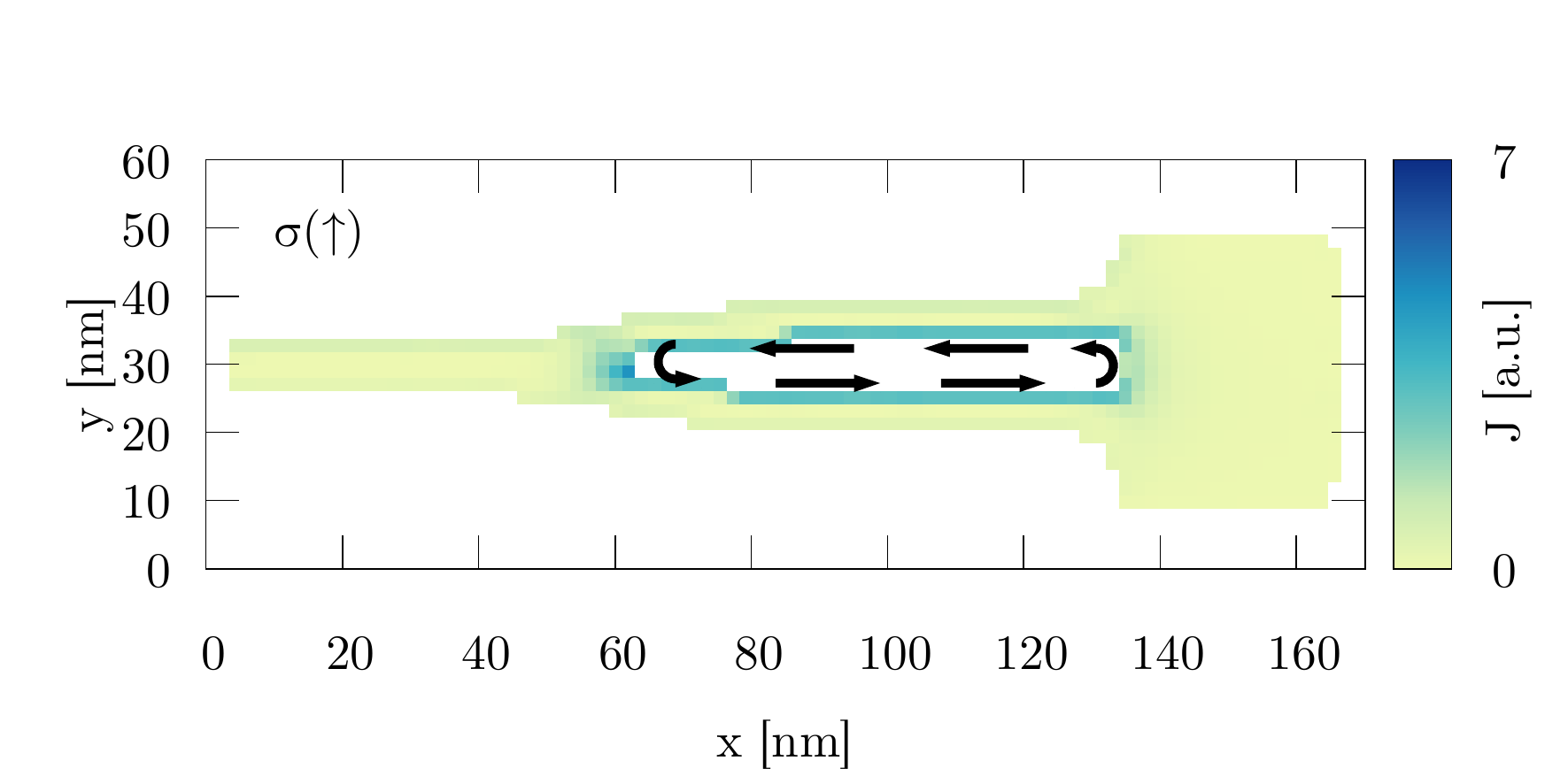}\put(-180,20){(b)}
\caption{The current map for $E_F = 1$ meV and $B=6$T (marked by dot in fig. \ref{fig:cutq}(c). Subplot (a) is for mode $k_i$ associated with spin down $[\sigma(\downarrow)]$ and (b) is for the spin up $[\sigma(\uparrow)]$. }
\label{fig:curr1}
\end{figure}

\begin{figure}[htbp]
\centering
\includegraphics[width=0.24\textwidth]{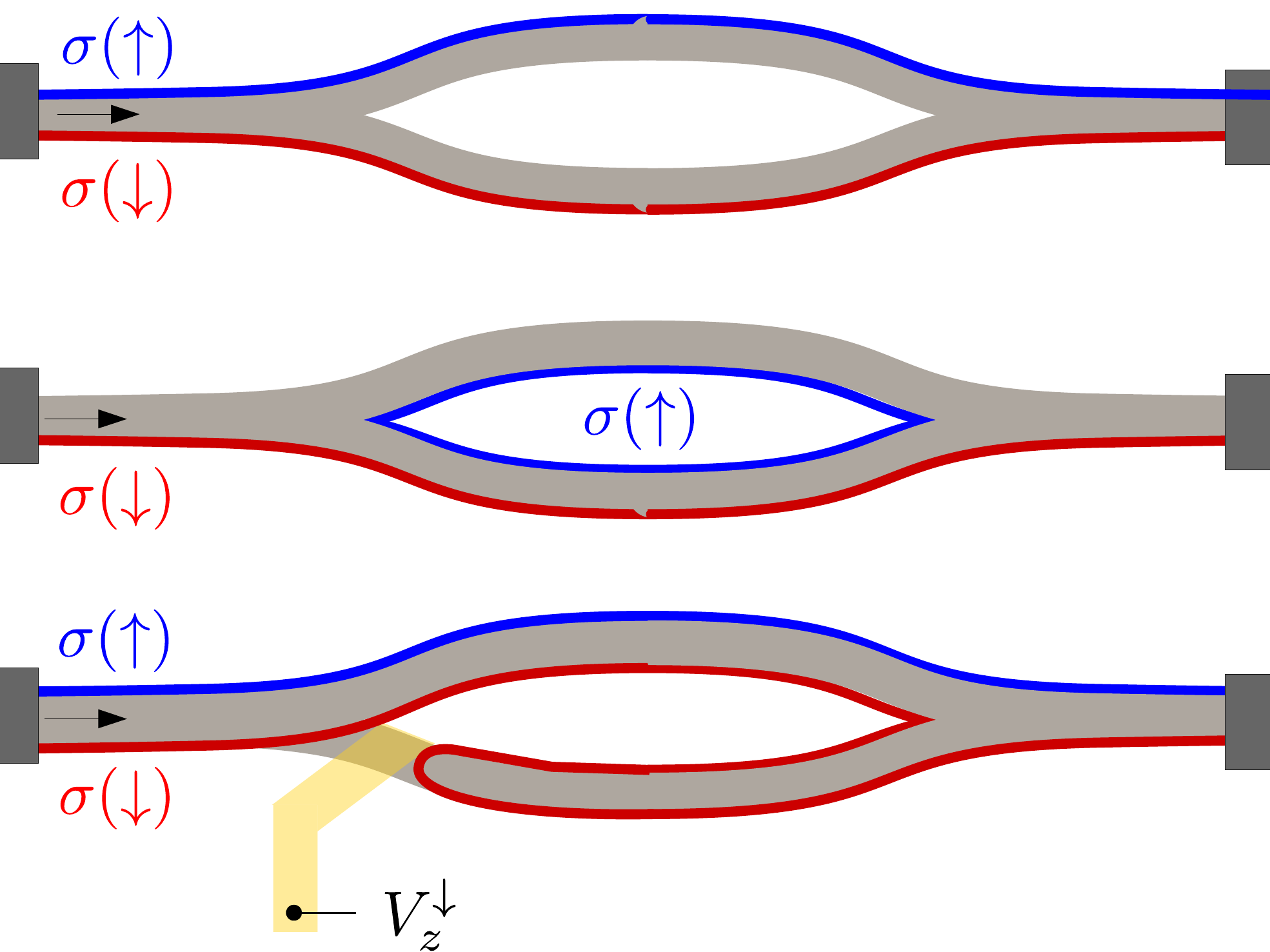}
\put(-2,90){(a)}\put(-2,60){(b)}\put(-2,30){(c)}
\caption{ Sketch of the quantum ring formed by reflection of
the fork channel of Fig. \ref{fig:sch1}.
 The current distributions observed in the QSH phase
are given for off-resonant (a) and resonant (b) conditions
and for the entrance of the lower lead closed  by a local electrostatic potential.
}
\label{fig:schr}
\end{figure}

In the QSH phase [Fig. \ref{fig:cutq}(b)] we
do not observe the regular AB oscillations 
even for both channels open. The wave function for each spin
passes through a single slit to the halfplane, so no Young
interference can occur. Instead, in Fig. \ref{fig:cutq}(b) we find  sharp peaks of conductance which correspond to localized resonances with the current circulation around the etched area [see Fig. \ref{fig:curr1}(b)
 for the point marked by the dot in Fig. \ref{fig:cutq}(b)].  The local electric field which
cuts off the upper or lower channels excludes the current circulation, and the rapid features of the conductance dependence on the external field disappear [Fig. \ref{fig:cutq}(a,c)].
  For one of the channels closed [Fig. \ref{fig:cutq}(a,c)]
the dominant spin in the detector is somewhat counterintuitive: when the lower channel --
preferred by the  spin-down currents -- is closed 
 the calculated $G_{dd}$ is  much larger than $G_{uu}$ [Fig. \ref{fig:cutq}(a)] for a general $B$.
The reason for this is that for $V_z^\downarrow \neq 0$ the spin-down current is directed to the upper channel, where it flows near its lower edge, thus closer to the detector.



\begin{figure}[htbp]
\centering
\includegraphics[width=0.45\textwidth,clip,trim=0cm 1.9cm 0cm 0cm]{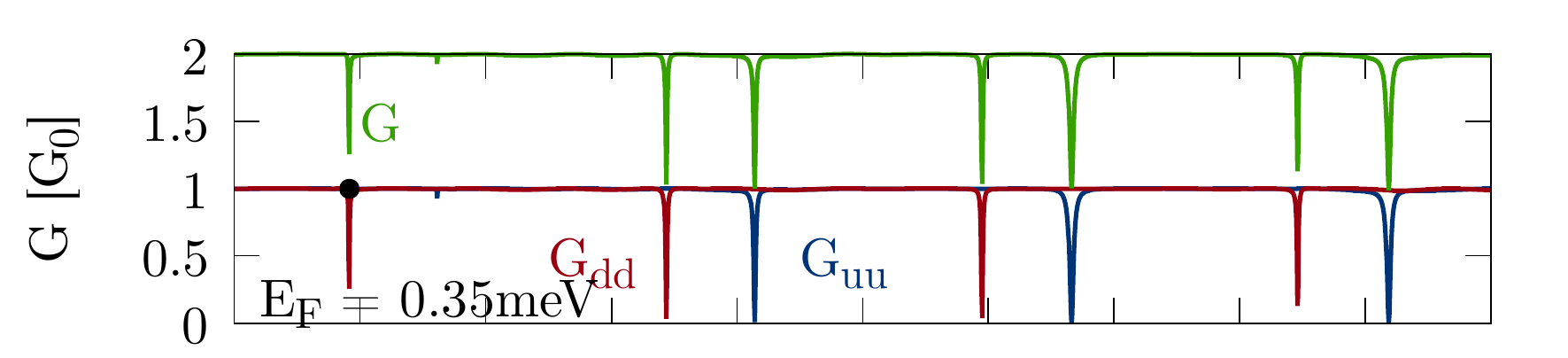}\put(-225,42){(a)}\\
\includegraphics[width=0.45\textwidth,clip,trim=0cm 1.9cm 0cm 0cm]{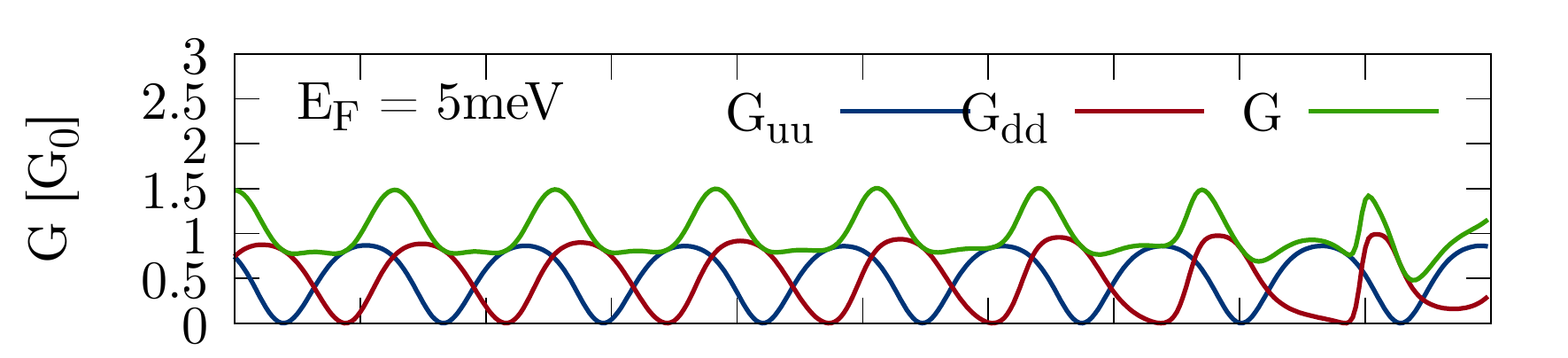}\put(-225,42){(b)}\\
\includegraphics[width=0.45\textwidth,clip,trim=0cm 0cm 0cm 0cm]{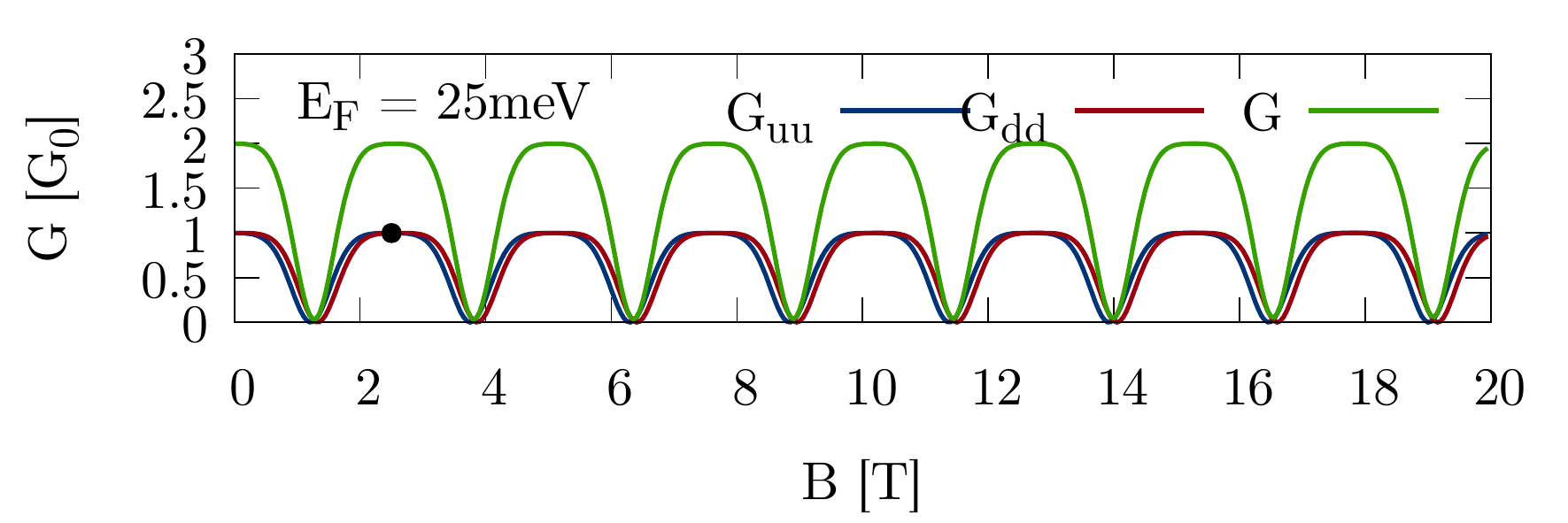}\put(-225,65){(c)}
\caption{ Conductance of the quantum ring in the (a) QSH regime $E_F=0.35$ meV, and
in the normal conditions (b,c)  for $E_F=5$ meV (b) and $25$ meV (c) in the absence
of vertical electric field.
}
\label{fig:circ_t}
\end{figure}

Similar control of the two-path interference effects can be obtained in a quantum ring [Fig. \ref{fig:schr}] formed
by reflection of the split channel of Fig. \ref{fig:sch1}(a). 
The calculated conductance in the QSH regime is given in Fig. \ref{fig:circ_t}(a) for $E_F=0.35$ meV.
For a general magnetic field 
the system is transparent for the electron flow [Fig. \ref{fig:schr}(a)].
Sharp dips of conductance appear [Fig. \ref{fig:circ_t}(a)] by interference with the localized loops of current stabilized near the inner edge of the ring [Fig. \ref{fig:schr}(b)].  Note that in the open system conductance peaked by interference with localized states, because signal received by detector came from leakage [Fig. \ref{fig:cutq}(b)] of the resonant current loop.
The localized resonances are only weakly coupled to the leads hence their long lifetime that is translated to narrow width of the resonances.  In the normal transport conditions [Fig. \ref{fig:circ_t}(b,c)] no sharp resonances appear and smooth AB
oscillations appear with the period independent of the energy. For higher energy [Fig. \ref{fig:circ_t}(c)]
the contribution to conductance of opposite spins become equal. 

\begin{figure}[htbp]
\centering
\includegraphics[width=0.45\textwidth,clip,trim=0cm 1.9cm 0cm 0cm]{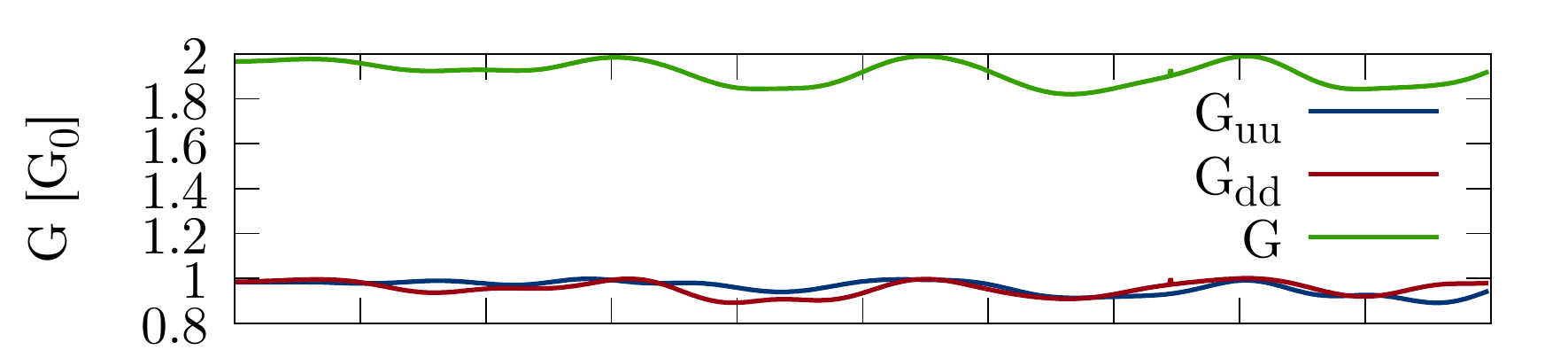}\put(-225,52){(a)} \put(-170,47){$E_F=0.35$ meV}\\
\includegraphics[width=0.45\textwidth,clip,trim=0cm 0cm 0cm 0cm]{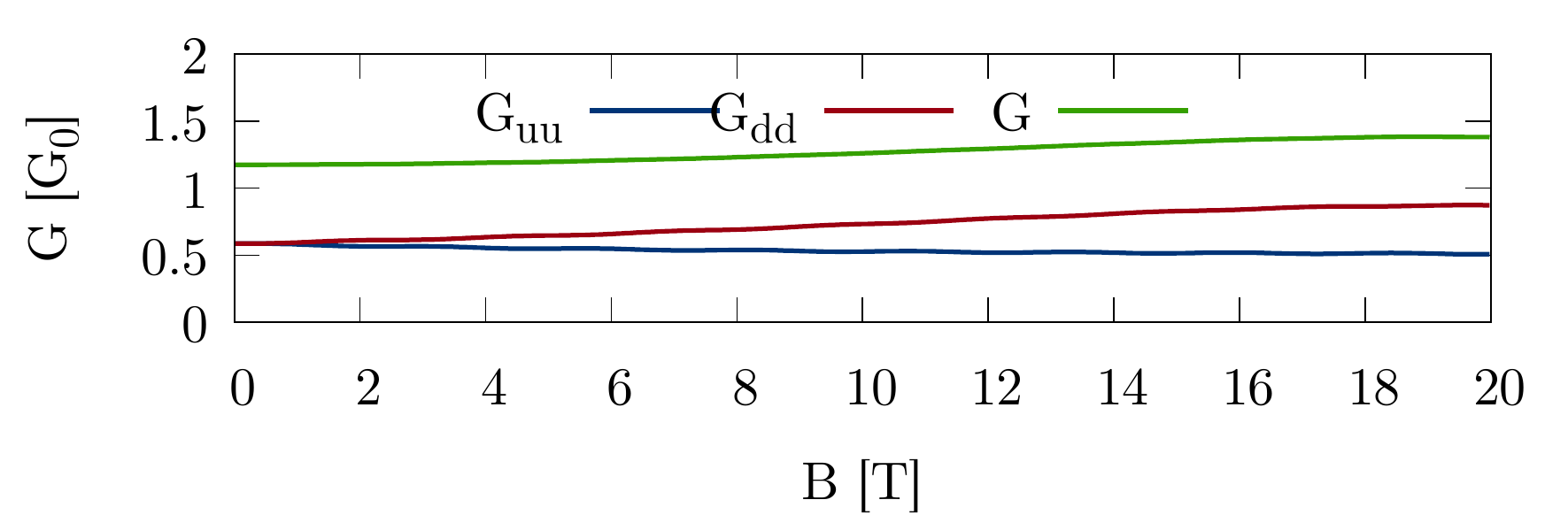}\put(-225,72){(b)} \put(-170,70){$E_F=5$ meV}\\
\caption{ (a,b) same as Fig. \ref{fig:circ_t}(a,b) only in presence of the local
electric field in the entrance to the lower lead [Fig. \ref{fig:schr}(c)].
}
\label{fig:circ_tf}
\end{figure}

For the gate that cuts off the current flow across  the lower channel [Fig. \ref{fig:schr}(c)]
the dips due to the localized states disappear in the QSH regime [Fig. \ref{fig:circ_tf}(a)]
and the AB oscillations are removed in the normal conditions [Fig. \ref{fig:circ_tf}(b)].
For the normal conditions at $B=0$ the closed lower channel reduces the conductance significantly. In the QSH conditions both spin currents find their way to the exit
of the ring, although the path for the spin down current becomes quite complex
[Fig. \ref{fig:schr}(c)].

In summary, we have demonstrated that  gated interference devices can be defined in silicene  to allow for detection of the quantum spin Hall transport conditions by reaction of the conductance  to the local electric fields closing one of the paths for the electron flow.  

\section*{Acknowledgments}
B.R. and A.M-K. are supported by Polish government budget for science in 2017-2021   within project "Diamentowy Grant" (Grant No. 0045/DIA/2017/46).
B.R. acknowledges the support of EU Project POWER.03.02.00-00-I004/16. B.S. acknowledges the support of NCN grant DEC-2016/23/B/ST3/00821. The calculations were performed on PL-Grid Infrastructure at ACK Cyfronet AGH.

\bibliographystyle{apsrev4-1}

\end{document}